# Behavioural analysis of interaction between individuals and a robot in the window

Federica Bertel

University of Turin, federica.bertel@edu.unito.it

Cristina Gena

Dept. of Computer Science, University of Turin, cristina.gena@unito.it

Matteo Nazzario, Irene Borgini

Intesa Sanpaolo Innovation Center, Turin, Italy, matteo.nazzario@intesasanpaolo.com, irene.borgini@intesasanpaolo.com

The aim of the current research is to analyse and discover, in a real context, behaviours, reactions and modes of interaction of social actors (people) with the humanoid robot Pepper. Indeed, we wanted to observe in a real, highly frequented context, the reactions and interactions of people with Pepper, placed in a shop window, through a systematic observation approach. The most interesting aspects of this research will be illustrated, bearing in mind that this is a preliminary analysis, therefore, not yet definitively concluded.

## 1 INTRODUCTION AND RELATED WORK

The *Socially Interactive Robots* [1] are robots that are able to establish a peer-to-peer relationship with humans (i.e. designed with a view to play a key role in social interactions with the latter) through a whole series of specific social characteristics, including, for example: *expressing and/or perceiving emotions; communicating with a high level of dialogue; learning/recognising behavioural patterns of other social agents; establishing/maintaining social relationships; using natural signals (gaze, gestures, etc.); displaying a distinctive personality and character; learning/developing social skills* [1].

They are also endowed with a specific flexibility, which allows them to adapt their behaviour and type of interaction according to the context by taking on different roles (e.g., partners, colleagues, assistants, etc.) with various types of users. Over time, experiments have been carried out on robots with the function of: personal assistants for the elderly, collaborators in human teams, welcoming figures to groups of various numbers of participants at public events (such as trade fairs and museum exhibitions), and educational toys or therapeutic tools (for example, in autism and diabetes therapies) [1].

These experiments have been very successful and have helped to examine what social processes are triggered when people encounter robots or spend time with them, as shown by a lot of researches considered by Horstmann and Krämer [2], which shows how the previously mentioned social and physical characteristics of robots influence users' perceptions by perceiving robots as partially alive, when in fact they are nothing more than a simulation of human social interactions [3] [4]. The resulting confusion and uncertainty lead users, especially in the first encounters, to gather information and try to understand this new unknown phenomenon in order to know what to expect [5]. This happens in two ways: on one hand, through the first impression they have of the robot [6], which can subsequently be confirmed or denied by throughout observations of the attitudes, values and behaviour of the robot as a whole (its personality) [7]. On the other hand, considering how uncommon it is for people to meet robots in person and deal with them [8], in [9] Sandoval et al. hypothesised that users try to fill this lack of information through the media, especially science fiction products, which would seem to influence their expectations accordingly. This would have a decisive impact not only on the acceptance of social robots themselves, but also on the future choice to use and include them in one's life [10].

Horstmann and Krämer [2] therefore tested, through two studies, how much people's expectations and preferences can be influenced by science fiction entertainment in the absence of a more direct and intimate contact with them.

In the case of respondents who had not previous live contact with social robots, despite the fact that they were willing and open to the idea of interacting with robots in the near future, they also expressed some concern about the robots' capabilities and skills, fearing that they might become dangerous. This was motivated by the authors [2], who observed that most of the respondents mentioned science fiction films during the interviews, assuming that these fears and concerns stem from negatively perceived memories of fictional robots.

According to [5], the lack of close and frequent interactions with social robots, in the absence of other reliable information, makes people more vulnerable, leading them to base their expectations on the first available sources of information (the mass media and science fiction products) rather than being completely unprepared, remaining within the framework of a predictable explanation for the new phenomenon.

In contrast, technophiles and those who have experienced previous interactions with social robots seem to be particularly inclined to have high expectations of their skills and abilities, less concerned about the potential dangers arising, and hopeful that they will be able to experience them much more closely in the near future [2]. This would confirm the hypotheses of other studies, according to which real contact between users and robots can reduce negative prejudices towards the latter, something that had already been found among human interaction partners [11].

Horstmann and Krämer 's observations [2] would seem to confirm that society and people are not yet used to seeing and experiencing robots on a daily basis, and that they know very little or only marginally about them. This is particularly true from the perspective of social robotics research, as the last decade has seen limited use of live experiments between humans and robots, as highlighted in [12]. From an overall analysis of 86 papers (published from 2008 to 2018), in [12] Lambert et al. verified how the number of user studies has seen some decline from 2010 to 2016, despite being present in *63%* of the papers. Specifically, *51%* reported only one user study, *37%* reported no experiment or no experiment results, *7%* reported two experiments, while *5%* reported three or more experiments.

The application areas that made most use of user studies were, in order, *Social effects* (which focuses on the evaluation of the effects of a social interaction with a robot, designed to provoke a specific reaction) with *31%*, *Companionship* with *24%*, *Social Definitions* (which focuses on the study of the expectations assumed towards robots in a social context, by users) with *22%*, *Education and Healthcare* in *11.5%*. In addition, the area with the highest number of users involved in the research was *Social Definitions* (79 users on average), as many of the studies considered were conducted through digital tools such as *Amazon Mechanical Turk* [12].

This general trend obviously has reasons. Until now, social robots were not as commercially widespread, they costed a lot of money and limiting therefore experiments to laboratory settings with small groups of people. However, the situation is changing, and if we want social robots to be more present in complex social contexts (those that people experience on a day-to-day basis), more live experimentation is needed [13]. Jung and Hinds, in [13], have reported several studies pointing out how proximity with a social robot profoundly changed the social structure, roles and rhythms within it, in a way that could never be observed in a laboratory setting.

Despite this, research in the field remains quite rare and is not sufficient on its own to create a solid theory that can refine the planning and design of social robots in these contexts. Some work is moving in this direction, trying to obtain valid design lines for future design through the application of qualitative, or hybrid (both qualitative and quantitative) methodologies. For example, in [14] and [15] the use of *Sequential Observational Analysis* has been particularly useful [16]. This approach studies, through field observations, the behaviour and activities of subjects engaged in activities and interactions alone or in groups, with the aim of gathering important information on the management of relationships, on changes and difficulties, and on the robot's ability or shortcomings in being able to engage users [14]. Through behavioural observation, the two studies were able to identify sequences of events/behaviours that enabled users to manage their interaction with the robots.

Also in the Sugar, Salt & Pepper project [17] [18], a research project focused on the use of the Pepper robot in a therapeutic laboratory on autonomy aimed at promoting functional acquisitions in highly functioning (Asperger) children with autism, the authors proposed observational approaches. From observation analysis and structured meeting notes (trainee students filled out evaluation forms provided by psychotherapists, noting the children autonomy's progress in a diary with the helping of rating scales), the authors realized that, on the one hand, the robot personality and its character should be better defined order to satisfy the curiosity and the expectations of children, and to make it more credible and engaging; on the other hand, the lack of autonomy of robot's dialogue together with the peculiarity of autistic functioning distorted the essence of the conversation. The exchange of utterances did not produce the pleasure of sharing but was functional to obtaining something more concrete, such as searching for information. The authors concluded that, if the goal is reaching a typical conversation, the results could always be unsatisfactory in this context.

In [14] Sabanovic et al. described the experience of GRACE, a mobile robot that was tasked with locating a team member wearing a pink hat during a conference in Pittsburgh. To find him, GRACE had to engage in brief social interactions with conference participants, asking them for information and directions and whether they had seen this person or not. The communication with the robot was managed through the touchscreen, where users could give the requested directions to the robot and all interactions were recorded through cameras placed on it. The observed behaviours were: spatial movements, gestures, and glances both when conference participant were directly engaged in the interaction with GRACE and when they did not [14].

These allowed the authors [14] to understand the extent to which the physical and social environment in which GRACE was immersed influenced her interactions, in a way that was unexpected and not contemplated in relation to how it was initially designed. For example, the characteristics of the spaces in which GRACE was placed had different effects on its ability to instigate curiosity and interaction on the part of people. In fact, GRACE was placed in three environments: the conference reception (a large room full of people), a corridor (where people moved from one conference to another) and a banquet (a small social event held in the corridor). In particular, high rates of engagement and positive attitude of people towards GRACE were recorded both in the reception and at the banquet, with longer interaction sessions. Conference participants were inclined to want to continue interacting with GRACE, even if it had disengaged from the conversation [14]. According to the authors [14], this could be due to the nature of the spaces, as in the corridor, due to its transient and passing nature, participants were equally likely to interact or not with the robot with shorter interaction sessions. Furthermore, Sabanovic et al. [14] observed, in the reception area and at the banquet, a correlation between the gaze movement of passers-by and the subsequent interaction with GRACE (gaze interaction was detected in 30% of the cases and in 82% resulted in subsequent interaction with the robot). The ability to detect the gaze of passers-by, would be particularly useful for robots, as it would allow them to engage more people [14]. Sabanovic et. al [14], also designed GRACE so that it could sustain interactions with one person at a time. Observations have shown, however, that there is an equal chance (especially in very social events full of people) that GRACE can interact with groups of people (two to several people at a time) as well as with single people. This was most frequently observed in the reception area. In 53% of the observations, GRACE interacted with multiple people who either collaborated or took turns to help it find its team member [14]. Since GRACE's interface is organised to receive input from only one person, conference participants organised themselves among themselves by discussing the robot, what directions to give it, usually with one person interacting directly on the interface and the other collaborating indirectly or waiting their turn to act in turn. This caused confusion in GRACE, with some interruptions during the interaction [14]. All this valuable information would allow to improve the design of the robot by making it more adaptive with respect to the reference environment (the conference) by decreasing unforeseen problems and malfunctions, but also increasing the possibilities of engagement with users [14]. For example, if GRACE was able to consider the density of the crowd in a room and their movements, it could manage interactions accordingly (in a crowded room, for example, it would understand that it would only have to interface with people to ask for space, whereas in a less crowded room it could achieve longer and more in-depth interactions) [14].

In [15], a similar work was done with middle school children who interacted with PLEO, a dinosaur toy robot. PLEO, has several sensors under the skin and limbs, speakers, and microphones, as well as a fairly developed personality capable of displaying different emotions (such as fear, happiness, curiosity) and drives (hunger, sleep). The children had previously experienced PLEO live, in their classroom. The aim of the second meeting, through the observation of the play sessions, was to understand how the children's previous knowledge of PLEO could influence the subsequent interaction (and thus, how the interaction and attitude of the users can change after a certain period spent with the robot) [15]. During the observation in the laboratory, the children were left free to play with it. Immediately afterwards, focus groups were carried out with them to collect their hot perspectives. During the observations, Andrés et al. [15] focused on the emotions elicited by PLEO and how the children interacted with him. Mainly positive and neutral emotions (mostly the latter) were recognised during the users' play sessions [15]. The duration of the sessions themselves was variable, as it was the children who decided how much time to spend with the robot (interactions lasting from about five to twenty minutes were observed). In this respect, an interesting detail was the children's maintenance of eye contact with the robot throughout the interaction, a sign that it gathered their attention, which was lacking towards the end when the children were either satisfied or bored [15]. Another interesting detail observed was the lack of physical distance between PLEO and the participants. Since the children were already familiar with the robot, they did not feel fear towards it, but held it in their arms the whole time, cuddling and caring for it and, above all, exploring it through touch [15]. According to the authors [15], these details may be useful in understanding how to manage and direct the design of social (specifically companion) robots in long-term interactions, considering that maintaining attention is a crucial variable if users are not to become bored too soon.

The involvement of children has also been a main pillar in the design of Wolly [19] [20], a social and affective educational robot. The authors involved kids as co-designer helping in shaping form and behavior of the robot. They used a co-design methodology and analyzed all the collected materials using the Grounded Theory approach, described in Section 2.1.1.

The aim of the current research is to analyse and discover in a real, highly frequented context, the interactions of people with the humanoid robot Pepper, *placed in a shop window*, through a systematic observation approach. Firstly, this work will analyse the research methodologies applied and their theoretical assumptions. Then, the context in which the research took place will be discussed, applying the same methodologies illustrated above and analysing the initial results obtained. Finally, the conclusions and some observations will be presented.

## 2 EXPERIMENTAL OBSERVATION IN REAL CONTEXT.

The Observational Analysis [16] was held in the Area X zone of Intesa Sanpaolo in Turin, Italy, and took place over a period of approximately two months from *September to October 2022*. Area X is an experiential zone dedicated to insurance protection, where users can create their own awareness through some dedicated virtual reality games [21]. The storefront where Pepper's is located overlooks a busy street and, on the other side of the street, there is a public transport stop, resulting in movement and passage by a varied target group (e.g., young people, adults, the elderly, children, families and even tourists etc.). Thus, Pepper was positioned in plain sight, where it can easily be noticed, as shown in Figure 1.

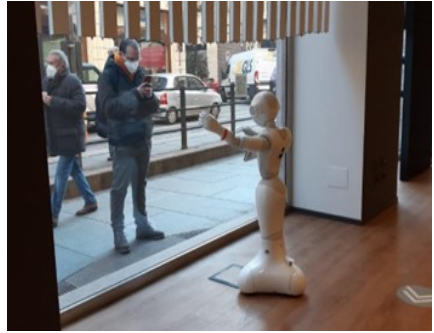

Figure 1 Robot Pepper in Area X

The observer who carried out the observational analysis was positioned at the back of the room, right behind Pepper. People can interact with the robot by scanning the QR code on the window with their mobile phone, then they will have to type a numeric code (PIN) appearing on the robot's tablet shortly afterwards. At that point, guided by Pepper's voice, they can decide from the menu on the phone what to have Pepper do.

The users can choose to listen to audio insights on Area X and the Innovation Centre, narrated by Pepper, or have it played by choosing one of these activities (even multiple times): Selfie, Swag, Dance and Exult. The interaction is programmed to last no longer than three minutes, at the end of which the application asks the user if he/she wants to continue playing or leave the session. If other people scan the QR code while Pepper is already engaged, the application will put them on 'hold'. In addition, Pepper has been programmed so that at times when it is not engaged in other interactions, it moves every few minutes, e.g., waving to passersby, making arm gestures[1]. Thanks to its sensors, Pepper can perceive movements around it and see who is approaching, being able to follow it with its eyes and head. For privacy concerns of passers-by, the cameras positioned in the eyes have been kept switched off. The described interaction was designed in order to propose a set of gamified activities, as already successfully experienced in [22] to make the experience more meaningful and enjoyable, and collect feedback in a real context of use.

---

[1] The application presented here was developed and designed by the Artificial Intelligence and Robotics Laboratory of Intesa SanPaolo Innovation Center S.P.A.

## 2.1 Observational analysis in Area X

Referring to this experimentation, both quantitative and qualitative data were found to be of particular interest. The qualitative data were obtained through Sequential Observational Analysis, which has been explained above, while the quantitative data were collected through Pepper the robot. Subsequently, a first part of the collected data was processed and interpreted through the procedures of both Sequential Observational Analysis [16] and Grounded Theory [24] which will be explored in more details in the following paragraphs.

It is important to remember that this work contains only provisional evidence as the interpretation and analysis of the data is not yet complete.

### 2.1.1    Grounded Theory Methodology.

Before moving on, this section will elaborate on the methodological procedure with which the analysis of the behavioural data collected in Area X was carried out.

Grounded Theory is based on an inductive methodological approach [24], in which the researcher can extrapolate information and develop hypotheses about the phenomenon from the analysis of the collected data by combining and interplaying different evaluation methodologies (e.g., quantitative and qualitative), as described and performed for instance in [23]. Grounded Theory, in fact, is based on philosophical and sociological assumptions, such as *Symbolic Interactionism* and *Pragmatism* [24]. These assumptions make this theory, therefore, an open and flexible methodology, allowing the researcher ample freedom in interpreting the data and identifying possible explanations of the phenomenon, without neglecting the validity of the data and carrying out the correct methodological procedure. The procedure underlying Grounded Theory is *coding*, an analytical process by which the researcher can interpret the collected data [24]. In order for this to be performed correctly, Corbin and Strauss [24] make it clear that the data collection phase cannot be separated from the coding phase, rather the two phases are interconnected processes. This implies that the observer, by coding the first data collected, can grasp a certain number of details and clues manifested with respect to the phenomenon observed, which can be confirmed or refuted with subsequent collections and interpretations, refining his or her understanding of the phenomenon.

There are three coding stages that will be discussed in a moment: Open Coding, Axial Coding and Selective Coding [24].

During the first phase, previously collected data are unpacked by identifying specific actions/interactions/events/behaviours, manifest clues to the observed phenomenon. These are compared with each other to identify similarities and differences. Those that are similar are collected and represented through a *conceptual label* (an abstract concept identifying the observed event/behaviour). Subsequently, the observer will collect similar conceptual labels into *categories*: a set of different actions/interactions/behaviours occurring in a certain situation. Let us take the case, for example, that the observer has analysed the interactions between nurses and patients and has identified several concepts, and upon further coding realises that some of these are directed towards providing comfort for the patient. *Comfort work* will be identified as a category [24].

The categories can be further refined by analysing their *properties* and *dimensions* (identifying different types of comfort work, duration, how they are performed, etc.) [24].

In the second phase, Axial Coding, the observer will further deepen and refine the identified categories. Using the previous example, the observer will delve from the data into the specific *conditions* under which *comfort work* occurred, in which *contexts* and with which events/behaviours and *consequences*. These are identified as *sub-categories* and are specifically and directly linked to a category. This allows the observer to identify all variations of the phenomenon under investigation [24].

In Selecting Coding, the observer moves on to the final stage of the search, which consists of identifying a *core category*, which is able to identify and synthesise the search phenomenon into a single concept. Not only that, the core category is related to all the categories and subcategories identified in the coding. This has a twofold positive aspect, which is the goal of Grounded Theory. The more *abstract* the core category (and the categories and concepts related to it), the more *generic* and thus applicable the identified theory will be. The latter is also particularly *specific*, as the phenomenon has been observed under specific conditions, in particular contexts with the occurrence of specific events/behaviours (the categories and concepts), thus making it *reproducible* [24].

At this still preliminary stage of the research, as it will be fully discussed in the next section, it was possible to identify the first conceptual labels. These were then described by means of a whole series of statistical measurements typical of Observational Analysis.

*2.1.2   Observational analysis in Area X: data collection and analysis.*

As mentioned earlier, during the Area X trial it was possible to collect both quantitative and qualitative data. In the former case, through the data collected by Pepper during interactions, dates and times of interaction, averages of QR code scans and completed interactions have been identified.

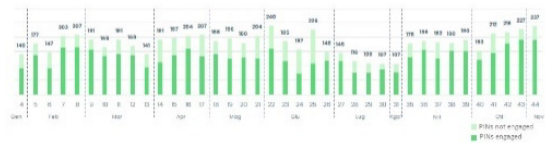

Figure 2 PINs generated per week in the period January - November 2022

In Figure 2, there is an example of the information the system collected since interactions with Pepper. In this case, the number of PINs generated, and the number of PINs used for interaction were analysed and compared for each week from 24$^{th}$ of January to 6$^{th}$ of November 2022, with a break from 7$^{th}$ to 28$^{th}$ of August when Pepper was switched off. What can be deduced is that approximately *72% of users* used the PIN to engage Pepper[2]. The observational activity took place in September and October and resulted in *183 observations* being recorded.

In Figure 3 are displayed, according to the data analysed by the system against the interactions made, the favourite activities chosen by users during the engagement[3] . First and foremost, there are dancing and selfies, and in general the gaming category.

---

[2] It must be specified that when the robot is already engaged in an interaction, the scanning of the QR code does not generate a PIN and the user is invited to try again after a certain amount of time. It is therefore not possible to determine the number of unsuccessful attempts to read the QR code when the robot is already engaged.

[3] The graph considers users' choices of activities as unique, i.e., they are counted only on the user's first selection. This means that if an activity or activities are repeated during the interaction, the system will ignore them.

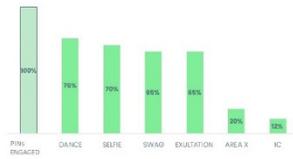

Figure 3 Preferred types of user engagements

About qualitative data, that is what was collected and deduced directly from the observations, it has been possible to identify the population groups involved in the interactions and some conceptual labels, with annexed frequency percentages of the total. As far as the sample involved is concerned, it consists of a diverse subjects, characterised by the passage mainly of groups (family members, friends, young and adult couples) and single passers-by of different ages for a grand total of *379 people involved*, with a slight female predominance.

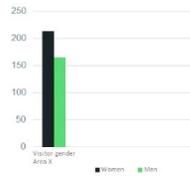

Figure 4 Visitor gender Area X

Although the total sample is 379 people, this does not imply that there was an equal number of scans. On the contrary, from the observations, hardly more than one scan per subject or per group was performed. In the latter case, one person oversaw scanning the QR code while they worked together to decide what to have the robot do. This resulted in 33 out of 150 scans not being performed, making a total of 183.

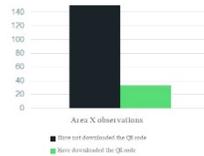

Figure 5 Scans and non-scans of QR code in Area X

Thanks to the systematic we analysed a whole sets of actions and behaviours that individuals performed during the interaction, both towards Pepper and with other people with them, both before and after the scan.

The recurring behaviours of people passing and/or stopping in front of Pepper's window, were observed, and collected under certain conceptual labels ("*Greeting Pepper*", "*Returning to Pepper*", "*Observing Pepper because they do not know what to do*" and "*Observing Pepper but are not interested in interacting*"), following the methodological approach of Grounded Theory seen above.

"*Greeting Pepper*" was a frequent behaviour in *51.67%* of our observations. People tend to greet the robot, either of their own accord or in response to the robot. *11% return to see it*, days or weeks later. *31% observe it wanting to interact but not understanding how*. Consequently, users focused on *attracting the robot's attention* by moving or waving. This was also confirmed by some passers-by's warm comments about how they had not noticed the QR codes at all or that *they had no idea how the robot worked*. Finally, in *58%* of the cases *the users merely observed it*, out of curiosity for a moment, *but had no interest in interacting* further with it demonstrated by the fact that while they were looking at it, they did not stop in their tracks.

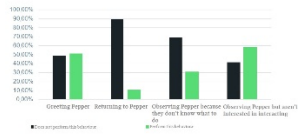

Figure 6  Some behaviours highlighted in Area X

Compared to these initial results, Pepper gained success among people, who also tend to *humanise it* (i.e., treating it like a person, even like a real child). Despite the desire to interact with it, the understanding of how to do so is not so immediate. It is likely that its lack of responsiveness also does not help to convince people past playing with him.

## 3  CONCLUSIONS

It has already been stated several times that the work presented here is still in its early stages, so there is still much to discover. Nevertheless, the first evidence in Area X, shows that the interaction with Pepper has been quite successful but that the interaction could and should be improved to attract more users.

From the examples and observations above, some *critical points* have already been identified. For instance, it should be easier to identify the QR codes in Area X since they represent the core of the interaction with the robot. Since the users' field of vision is completely focused on it, a good option would be to make more use of the tablet (perhaps by inserting a more specific message that could guide the user on what to do or by inserting the QR code directly there). It is important that Pepper leads the interaction at the beginning, making the first move so that the user understands how he can behave and what to do.

The next step is to complete the analysis phase of the collected observations by proceeding with further labelling, then reaching the goal of the experimental thesis: to *develop a theory* capable of answering the research questions posed at the beginning (how do people behave with social robots and why?).

This will help developing *design guidelines* to ensure more effective, efficient, and satisfying human-robot interactions.

## ACKNOWLEDGMENTS

This work has been partly funded by Intesa Sanpaolo Innovation Center, Turin, Italy.

## REFERENCES

[1]  Terrence Fong, Illah Nourbakhsh and Kerstin Dautenhahn. 2003. A Survey of Socially Interactive Robots: Concepts, Design and Applications. Robotics and autonomous systems*,* 42, 3-4(2003), 143-166.

[2]  Aike C. Horstmann and Nicole C. Krämer. 2019. Great Expectations? Relation of Previous Experiences with Social Robots in Real Life or in the Media and Expectancies Based on Qualitative and Quantitative Assessment. Frontiers in Psychology Sec. Human-Media Interaction*,* 10,939(2019). doi: 10.3389/fpsyg.2019.00939

[3]  Byron Reeves and Clifford Nass. 1996. The Media Equation: How People Treat Computers, Television and New Media Like Real People and Places. Stanford, CA: CSLI Publications.

[4] Jonathan Mumm and Bilge Mutlu. 2011. Human-robot proxemics: physical and psychological distancing in human-robot interaction. In HRI '11: Proceedings of the 6th international conference on Human-robot interaction. Lausanne, Switzerland, 331-338. https://doi.org/10.1145/1957656.1957786

[5] Charles R. Berger and Richard J. Calabrese. 1975. Some Explorations in Initial Interaction and Beyond: Toward a Developmental Theory of Interpersonal Communication. Human Communication Research, 1,2(1975 December), 99–112.

[6] Solomon E. Asch. 1946. Forming impressions of personality. The Journal of Abnormal and Social Psychology, 41,3(1946), 258–290.

[7] Susan M. Andersen and Roberta L. Klatzky. 1987. Traits and social stereotypes: Levels of categorization in person perception. Journal of Personality and Social Psychology, 53,2(1987), 235–246. https://doi.org/10.1037/0022-3514.53.2.235

[8] R. Van Oers and E. Wesselmann. 2016. Social Robotics. Amstelveen: KPMG Advisory N.V.

[9] Eduardo B. Sandoval, Omar Mubin and Mohammad Obaid. 2014. Human Robot Interaction and Fiction: A Contradiction. In: Beetz, M., Johnston, B., Williams, MA. (eds) Social Robotics. ICSR 2014. Lecture Notes in Computer Science Sydney, NSW, Australia, 54-63. Springer International Publishing. https://doi.org/10.1007/978-3-319-11973-1_6

[10] Viswanath Venkatesh and Fred D. Davis. 2000. A Theoretical Extension of the Technology Acceptance Model: Four Longitudinal Field Studies. Management Science, 46,2(2000),186-204. https://doi.org/10.1287/mnsc.46.2.186.11926

[11] Jens Binder, Hanna Zagefka, Rupert Brown, Funke Friedrich, Thomas Kessler, Amelie Mummendey, Annemie Maquil, Stephanie Demoulin and Jacques-Philippe Leyens. 2009. Does contact reduce prejudice or does prejudice reduce contact? A longitudinal test of the contact hypothesis among majority and minority groups in three european countries. Journal of Personality and Social Psychology, 96,4(2009),843–856. https://doi.org/10.1037/a0013470

[12] Alexis Lambert, Nahal Nourozi, Gerd Bruder and Gregory Welch. A Systematic Review of Ten Years of Research on Human Interaction with Social Robots. International Journal of Human–Computer Interaction,36,19(2020), 1804-1817. https://doi.org/10.1080/10447318.2020.1801172

[13] Malte Jung and Pamela Hinds. 2018. Robots in the Wild: A Time for More Robust Theories of Human-Robot Interaction. ACM Transactions on Human-Robot Interaction,7,1(2018 May),1-5. https://doi.org/10.1145/3208975

[14] Selma Sabanovic, Marek P. Michalowski and Reid Simmons. 2006. Robots in the wild: Observing human-robot social interaction outside the lab. In 9th IEEE International Workshop on Advanced Motion Control, 2006, IEEE. Istabul, Turkey, 596-601. 10.1109/AMC.2006.1631758

[15] Amara Andrés, Diego E. Pardo, Marta Diaz and Cecilio Angulo. 2015. New instrumentation for human robot interaction assessment based on observational methods. Journal of Ambient Intelligence and Smart Environments, 7,4(2015), 397-413. 10.3233/AIS-150331

[16] Roger Bakeman and John M. Gottman. 1997. Observing interaction: An introduction to sequential analysis. Cambridge university press.

[17] Cristina Gena, Claudio Mattutino, Andrea Maieli, Elisabetta Miraglio, Giulia Ricciardiello, Rossana Damiano, and Alessandro Mazzei. 2021. Autistic Children's Mental Model of a Humanoid Robot. In Adjunct Proceedings of the 29th ACM Conference on User Modeling, Adaptation and Personalization (UMAP '21). Association for Computing Machinery, New York, NY, USA, 128–129. https://doi.org/10.1145/3450614.3463422

[18] Cristina Gena, Rossana Damiano, Claudio Mattutino, Alessandro Mazzei, Andrea Meirone, Loredana Mazzotta, Matteo Nazzario, Valeria Ricci, Stefania Brighenti, Federica Liscio, Francesco Petriglia: "Preliminary results of a therapeutic lab for promoting autonomies in autistic children." *arXiv preprint arXiv:2305.02982* (2023)

[19] V. Cietto, C. Gena, I. Lombardi, C. Mattutino and C. Vaudano, "Co-designing with kids an educational robot," 2018 IEEE Workshop on Advanced Robotics and its Social Impacts (ARSO), Genova, Italy, 2018, pp. 139-140, doi: 10.1109/ARSO.2018.8625810.

[20] Gena, C., Mattutino, C., Perosino, G., Trainito, M., Vaudano, C., & Cellie, D. (2020, May). Design and development of a social, educational and affective robot. In *2020 IEEE Conference on Evolving and Adaptive Intelligent Systems (EAIS)* (pp. 1-8).

[21] Area X. Intesa SanPaolo Assicura. Retrieved November 30, 2022 from: https://www.areax.info/

[22] Amon Rapp, Federica Cena, Cristina Gena, Alessandro Marcengo & Luca Console (2016) Using game mechanics for field evaluation of prototype social applications: a novel methodology, Behaviour & Information Technology, 35:3, 184-195.

[23] Gena, C., Cena, F., Vernero, F., Grillo, P. The evaluation of a social adaptive website for cultural events. *User Model User-Adap Inter* **23**, 89–137 (2013) https://doi.org/10.1007/s11257-012-9129-9

[24] Juliet M. Corbin and Anselm Strauss. 1990. Grounded theory research: Procedures, canons, and evaluative criteria. Qualitative sociology, 13,1(1990), 3-21.